\newtheorem{thm}{Theorem}
\newtheorem{lem}{Lemma}
\newtheorem{prop}{Proposition}
\newtheorem{defn}{Definition}
\newtheorem{rem}{Remark}
\newtheorem{cor}{Corollary}

\documentclass[12pt]{iopart}
\usepackage{amssymb}
\usepackage{latexsym}

\def\dR{{\mathbb R}}

   \def\cD{{\mathcal D}}   \def\cH{{\mathcal H}}

\def\cP{{\mathcal P}}    \def\cT{{\mathcal T}}

\def\dom{{\rm dom\,}}

\def\PT{{\mathcal P}{\mathcal T}}
\newcommand{\Skindef}{[\raisebox{0.5 ex}{.},\raisebox{0.5 ex}{.}]}
\newcommand{\Skdef}{(\raisebox{0.5 ex}{.},\raisebox{0.5 ex}{.})}
\newcommand{\Norm}{\|\,\raisebox{0.5 ex}{.}\,\|}
\newcommand{\lln}{\lambda_{0}}
\newcommand{\qed}{{\hspace*{\fill}$\square$}\vspace{2ex}}
\def\mmm{{{\rm max}}}

\begin{document}

\title{On Domains of  $\PT$ Symmetric Operators Related to
$-y^{\prime \prime}(x) + (-1)^nx^{2n}y(x)$}

\author{Tomas Ya Azizov$^1$\footnote{The research of Tomas Ya.\ Azizov
is  supported by the Russian Foundation for Basic Researches (grant
08-01-00566-a).} and Carsten Trunk$^2$}
\address{$^1$ Department of Mathematics, Voronezh State
University, Universitetskaya pl.~1, 394006 Voronezh,
 Russia}

\address{$^2$ Institut f\"{u}r Mathematik, Technische Universit\"{a}t
 Ilmenau,  Postfach 10 05 65, D-98684~Ilmenau, Germany}
\eads{\mailto{azizov@math.vsu.ru},
\mailto{carsten.trunk@.tu-ilmenau.de}}

\begin{abstract}
In the recent years a generalization of Hermiticity was investigated
using a complex deformation $H=p^2 +x^2(ix)^\epsilon$ of the
harmonic oscillator Hamiltonian, where $\epsilon$ is a real
parameter. These complex Hamiltonians, possessing $\PT$ symmetry
(the product of parity and time reversal),  can have real spectrum.
We will consider the most simple case:  $\epsilon$ even. In this
paper we describe all  self-adjoint (Hermitian) and at the same time
 $\PT$ symmetric operators  associated to $H=p^2 +x^2(ix)^\epsilon$.
 Surprisingly it turns out that there are a large class of self-adjoint
 operators  associated to $H=p^2 +x^2(ix)^\epsilon$ which are not
 $\PT$ symmetric.
\end{abstract}





 \maketitle

\section{Introduction}

In the well-known paper from 1998 \cite{BB98} C.M.\ Bender and S.\
Boettcher considered the following Hamiltonians $\tau_{\epsilon}$,
\begin{equation}\label{Foundation}
\tau_{\epsilon} (y) (x) := - y^{\prime \prime}(x) + x^2
(ix)^{\epsilon} y(x), \quad {\epsilon} >0, \quad x\in \mathbb R.
\end{equation}
These complex Hamiltonians, possessing $\PT$ symmetry (the product
of parity and time reversal),  can have real spectrum. This gave
rise to a mathematically consistent complex extension of
conventional quantum mechanics  into $\PT$ quantum mechanics, see
e.g.\ the review paper \cite{Bender} and references therein. During
the past ten years $\PT$ models have been analyzed intensively,
e.g., Bethe Ansatz techniques were considered in \cite{DDT}, various
global approaches based on the extension of the above operators into
the complex plane are presented in \cite{BBM99,BCDM,S,Z}, $\PT$
symmetric perturbations of Hermitian operators can be found in
\cite{AMS,CCG,CCG2,CGS05}, extension theory for singular
perturbations of  $\PT$ symmetric operators in \cite{AGK,AK} and
considerations on spectral degeneracies in \cite{CGS,GGKN,GRS,Z1}.
In \cite{M1} $\PT$ symmetry was embedded in a general mathematical
context: pseudo-Hermiticity or, what is the same, the study of
self-adjoint operators in a Krein space, see also
\cite{AK1,J02,GSZ,LT04,T06,T06a}.

Usually, see, e.g., \cite{Bender,BB98,BBM99}, a closed densely
defined  operator $H$ in the Hilbert space $L^2(\dR)$ is called
$\PT$ symmetric if $H$ commutes with $\PT$. For unbounded operators
this is also a condition on the domains. It is the aim of this paper
to specify  $\PT$ symmetric operators connected with the
differential expression $\tau_{\epsilon}$ in (\ref{Foundation}).

Here we will restrict ourselves to the most simple case: We will
consider the differential  $\tau_\epsilon$ only in the case of
$\epsilon$ even. Hence, the above differential expression
$\tau_{\epsilon}$ in (\ref{Foundation}) will be either of the form
\begin{equation*}
\tau_{4n} (y) (x) := - y^{\prime \prime}(x) + x^{4n+2} y(x), \quad
{\epsilon} >0, \quad x\in \mathbb R.
\end{equation*}
if $\epsilon=4n$, $n\in \mathbb N$, or it will be of the form
\begin{equation*}
\tau_{4n+2} (y) (x) := - y^{\prime \prime}(x) - x^{4n+4}
 y(x), \quad {\epsilon} >0, \quad x\in \mathbb R.
\end{equation*}
in case $\epsilon=4n+2$.

We will describe all domains giving rise to a self-adjoint
(Hermitian) operator in $L^2(\dR)$ associated to $\tau_{\epsilon}$
which is at the same time $\PT$ symmetric.  This seems to be a
natural question. To our knowledge it is not addressed in earlier
publications.

Obviously, different domains have dramatic influence on the spectrum
of the corresponding operators. As an example, let us consider as a
possible domain the set $\widetilde D$ of all locally absolutely
continuous
 functions $f$ on the real line with a locally absolutely continuous
 derivative $f^\prime$  such that $f$ decays exponentially as $|x| \to
 \infty$. Define for $k\in \mathbb N$ the numbers $\alpha_k :=
 (4n+5-k)k^{-4n-6}e^k$ and $\beta_k :=
 (4n+6-k)k^{-4n-5}e^k$ and a function $c$, twice continuously
 differentiable on $[-1,1]$, such that the function $y_k$,
 $$
y_k(x) := \left\{ \begin{array}{cl}
(-\alpha_k x + \beta_k)e^x & \mbox{\rm if } x\leq -k,\\
(-x)^{-4n-5} & \mbox{\rm if }-k< x< -1,\\
c(x) & \mbox{\rm if }-1\leq x\leq 1,\\
x^{-4n-5} & \mbox{\rm if } 1< x< k,\\
(\alpha_k x + \beta_k)e^{-x} & \mbox{\rm if } x \geq k,
\end{array}\right.
 $$
is in $\widetilde D$. Obviously $(y_k)$ converges in $L^2(\dR)$ to
the function $y$,
$$
y(x) := \left\{ \begin{array}{cl}
(-x)^{-4n-5} & \mbox{\rm if } x< -1,\\
c(x) & \mbox{\rm if }-1\leq x\leq 1,\\
x^{-4n-5} & \mbox{\rm if } 1< x,
\end{array}\right.
 $$
which is not in $\widetilde D$. Moreover, $\tau_{4n+2}(y)$  is in
$L^2(\dR)$ and $(\tau_{4n+2}(y_k))$  converges in $L^2(\dR)$ to
$\tau_{4n+2}(y)$. This shows the following.
\begin{rem}
The densely defined operator $H$ defined via $ \dom H := \widetilde
D$,   $Hy := \tau_{4n+2}(y)$ for  $f \in \dom H$,  is not a closed
operator in $L^2(\dR)$. Hence, its spectrum covers the complex
plane,
$
\sigma(H) = \mathbb C.
$
\end{rem}

The domain which is naturally associated to $\tau_{4n}$ is the
maximal domain  $\cD_{\mmm}$. This is the set  of all locally
absolutely continuous
 functions $f\in L^2(\dR) $ with a locally absolutely continuous
 derivative $f^\prime$  such that $\tau_{4n}(f) \in L^2(\dR)$.
As $\tau_{4n}$ is in limit point case at $+\infty$ and $-\infty$, it
turns out, that there is only one self adjoint operator connected to
$\tau_{4n}$ which is also $\PT$ symmetric.

The more interesting case is $\epsilon=4n+2$. The differential
expression  $\tau_{4n+2}$  is in limit circle case at $+\infty$ and
$-\infty$ and it admits many different self-adjoint extensions.
These self-adjoint extensions are described via restrictions of the
maximal domain  $\cD_{\mmm}$ by ``boundary conditions at $+\infty$
and $-\infty$'' which determines the set of all domains of
self-adjoint extensions associated to
 $\epsilon=4n+2$. However, as a main result of this paper we
 characterize precisely which of these ``boundary conditions at $+\infty$
and $-\infty$'' give rise to  $\PT$ symmetric extensions. It turns
out, see Section \ref{section4} below, that surprisingly only a
rather small class of boundary conditions gives rise to
 $\PT$ symmetric extensions.
Hence, in order to obtain a $\PT$ symmetric operator associated with
$\tau_{4n+2}$ special attention has to be given to the right
boundary conditions.

 Limit point/limit circle classifications
 are a standard tool in Sturm-Liouville theory, we mention here only
 \cite{LS,W,W03,Z05}. Different  boundary conditions
 at $+\infty$ and $-\infty$ change the point spectra, a fact, which
 has to be taken into account for numerical simulations.

 All self-adjoint operators associated to  $\tau_{4n}$ and  $\tau_{4n+2}$
 share one common property: They commute also with the parity
 operator $\mathcal P$, hence they are also self-adjoint in a Krein
 space where the inner product is given by
 \begin{equation*}
[f,g]:=\int_\dR f(x)\overline{(\cP g)(x)}\,dx = \int_\dR
f(x)\overline{g(-x)}\,dx, \quad f,g\in L^2(\dR).
\end{equation*}
We describe the sign type properties of all extensions. This will
serve as a basis for the application of the perturbation theory in
Krein spaces which will be used in the study of the cases $\epsilon$
not even in a subsequent paper. A short introduction to self-adjoint
operators in Krein spaces is given in the next section.

\section{$\PT$ symmetric operators as self-adjoint
operators in Krein spaces}\label{section2a}

Recall that a complex linear space $\cH$ with a hermitian
nondegenerate sesquilinear form $\Skindef$ is called a {\em Krein
space} if there exists a
 so called {\it fundamental decomposition} (cf.\ \cite{AI,B,Krein})
\begin{equation}\label{decomp}
 \cH = \cH_+ \oplus \cH_-
\end{equation}
with subspaces $\cH_\pm$ being orthogonal to each other with respect
to $\Skindef$ such that $(\cH_\pm, \pm\Skindef)$ are Hilbert spaces.
Then
\begin{equation}\label{decompHilbert}
(x,x):=[x_+,x_+] -  [x_-,x_-], \quad x=x_++x_- \in\mathcal H \quad
\mbox{with } x_\pm \in \cH_\pm,
\end{equation}
 is an inner product and $({\mathcal H},\Skdef)$ is a Hilbert space.
All topological notions are understood with respect to some Hilbert
space norm $\Norm$ on ${\cal H}$ such that $\Skindef$ is
$\Norm$-continuous. Any two such norms are equivalent, see
\cite[Proposition I.1.2]{L}.
 Denote by $P_+$ and $P_-$ the
orthogonal projections onto $\cH_+$ and $\cH_-$, respectively. The
operator $J:= P_+-P_-$ is called the {\it fundamental symmetry}
corresponding to the decomposition (\ref{decomp}).

An element $x$  in a Krein space  $({\mathcal H},\Skindef)$ is
called {\it positive} ({\it negative, neutral}, respectively) if
$[x,x] > 0$ ($[x,x] < 0$, $[x,x] = 0$, respectively). For the basic
theory of Krein space and operators acting therein we refer to
\cite{AI,B} and, in the context of ${\mathcal PT}$ symmetry, we
refer to \cite{LT04}.

Let $A$ be a closed, densely defined operator  in the Krein space
$({\mathcal H},\Skindef)$. The adjoint $A^+$ of $A$ in the Krein
space $({\mathcal H},\Skindef)$ is defined with respect to the
indefinite inner product $\Skindef$, that is, its domain $\dom A^+$
is the set of all $x\in {\mathcal H}$ for which there exists a $z
\in {\mathcal H}$ with
$$
[Ay,x] = [y,z]\quad \mbox{for all } y \in \dom A
$$
and for these $x$ we put $A^+x:= z$. It is easily seen that (see,
e.g., \cite{L65,L})
\begin{equation}\label{DefA+}
A^+ = JA^*J,
\end{equation}
where $A^*$ denotes the adjoint with respect to the Hilbert space
inner product (\ref{decompHilbert}) and $J$ is the fundamental
symmetry corresponding to the decomposition (\ref{decomp}). The
operator $A$ is called self-adjoint in the Krein space $({\mathcal
H},\Skindef)$ if $A=A^+$.

The indefiniteness of the scalar product $\Skindef$ on $\cH$ induces
a natural classification of isolated real eigenvalues: A real
isolated eigenvalue $\lambda_0$ of $A$ is called of {\em positive}
({\em negative})
 {\em type} if all the corresponding eigenvectors are positive
(negative, respectively). It is usual to call such points of
positive type (negative type, respectively), see
\cite{AJT,ABJT,BPT,LMM1,L,LMaM} and in this case we write
$$
\lambda_0 \in \sigma_{++}(A) \quad ( \mbox{resp. }\lambda_0 \in
\sigma_{--}(A)).
$$
Observe that there is no Jordan chain of length greater than one
which corresponds to a eigenvalue of $A$ of positive type (or of
negative type). This classification of real isolated eigenvalues is
used frequently, we mention here only
\cite{BBJ1,BBJ2,CGS05,GK,GSZ,LT04}.

By $L^2(\dR)$ we denote the space of all equivalence classes of
measurable functions $f$ defined on $\dR$ for which $\int_\dR\vert
f(x)\vert^2 dx$ is finite. We equip $L^2(\dR)$ with the
 usual Hilbert scalar product
\begin{equation*}
(f,g):=\int_\dR f(x)\overline{g(x)}\,dx,\quad f,g\in L^2(\dR).
\end{equation*}
and we define
\begin{equation}\label{PTDef}
(\cP f)(x) = f(-x)\quad \mbox{and} \quad (\cT f)(x) =
\overline{f(x)}, \quad f\in L^2(\dR).
\end{equation}
Then $\cP^2 = \cT^2 = (\PT)^2= I$ and $\PT = \cT\cP$.  The operator
$\cP$ represents parity reflection and the operator $\cT$ represents
time reversal. Observe that the operator $\cT$ is nonlinear.

Usually, see, e.g., \cite{Bender,BB98,BBM99}, a closed  operator $H$
is called $\PT$ symmetric if $H$ commutes with $\PT$. For unbounded
operators this is also a condition on the domains. Therefore we will
repeat the notion of $\PT$ symmetry in the following definition
(see, e.g., \cite{Bender,CGS05,CCG}). We denote by $\dom H$ the
domain of the operator $H$.

\begin{defn} \label{PTDEF}
A closed densely defined  operator $H$ in $L^2(\dR)$
 is said to be $\PT$ symmetric if for all $f \in \dom H$ we have
$$
\PT f \in \dom H \quad \mbox{and} \quad \PT Hf = H\PT f.
$$
\end{defn}
Obviously, it follows from Definition \ref{PTDEF}
 $$
 \dom H = \dom H \PT.
 $$

To investigate the property of $\PT$ symmetric operators we will
need in the following the next lemma.

\begin{lem} \label{PTLEM}
Let $H$ be a closed densely defined  operator $H$ in $L^2(\dR)$ and
assume $\cT \dom H \subset \dom H$. The operator $H$  is  $\PT$
symmetric if and only if
$$
\cP\dom H \subset \dom H \quad \mbox{and} \quad \PT Hf = H\PT f
\quad \mbox{for all } f \in \dom H.
$$
\end{lem}

{\bf Proof.}\\
Let $f \in \dom H$. Let $H$ be $\PT$ symmetric. By assumption we
have $\cT f \in \dom H$ and, from the $\PT$ symmetry we conclude
$\PT \cT f = \cP f$ is in $\dom H$.

Contrary, for  $f \in \dom H$ we have by assumption $\cT f \in \dom
H$ and, hence, $\PT f \in \dom H$, that is, $H$ is $\PT$ symmetric.
\qed

The operator $\cP$ introduced in (\ref{PTDef}) gives in a natural
way rise to an indefinite inner product $\Skindef$ which will play
an important role in the following. We equip $L^2(\dR)$ with the
indefinite inner product
\begin{equation}\label{Pindef}
[f,g]:=\int_\dR f(x)\overline{(\cP g)(x)}\,dx = \int_\dR
f(x)\overline{g(-x)}\,dx, \quad f,g\in L^2(\dR).
\end{equation}

With respect to this inner product, $L^2(\dR)$ becomes a Krein
space. Observe that in this case the operator $\cP$ serves as a
fundamental symmetry in the Krein space $L^2(\dR), \Skindef)$. In
the situation where $\Skindef$ is given as in (\ref{Pindef}), it is
easy to see that as the positive component $\cH_+$ in a
decomposition (\ref{decomp}) the set of even functions, and as the
negative component $\cH_-$ the set of all odd functions of
 $L^2(\dR)$ can be chosen.

\begin{lem}\label{sa}
Let $H$ be a self-adjoint  operator $H$ in the Hilbert space
$L^2(\dR)$, $H=H^*$,  and assume that $H$ commutes with $\cP$. Then
$H$ is selfadjoint in the  Krein space $(L^2(\dR),\Skindef)$.
\end{lem}

The proof of this lemma follows immediately from (\ref{DefA+}) and
$H\cP = \cP H$. We mention that such operators are called {\it
fundamental reducible}, see, e.g., \cite{J3} and that they possess a
well developed spectral and perturbation theory, cf.\
\cite{ABJT,BJ,J3,J91,J98,Krein,LMaM,V1,V2}.

\section{Domains of $\PT$ symmetric operators in the case $\epsilon = 4n$}\label{section3}

We discuss first the more easy case $\epsilon = 4n$ for some $n \in
\mathbb N$, that is, we consider $\tau_{4n}$ defined according to
(\ref{Foundation}) via
\begin{equation*}\label{Foundation2}
\tau_{4n} (y) (x) := - y^{\prime \prime}(x) + x^{4n+2} y(x), \quad
x\in \mathbb R.
\end{equation*}
To this differential expression we will associate an operator $H$
defined on the maximal domain, i.e.,
$$
\cD_{\mmm}:= \{y \in L^2(\dR) : y,y^\prime \in AC_{loc}(\dR),
 \tau_{4n}y \in L^2(\dR)\},
$$
via
$$
\dom H := \cD_{\mmm}, \quad  Hy := \tau_{4n}(y) \quad \mbox{for } f
\in \dom H.
$$
Here and in the following $AC_{loc}(\dR)$ denotes the space of all
complex valued functions which are
 absolutely continuous on all compact
 subsets of $\dR$.

In the following theorem we collect some of the properties of $H$.
Recall that the differential expression $\tau_{4n}$ is called in
limit circle at $\infty$ (at $-\infty$) if all solutions of the
equation $\tau_{4n}(y) - \lambda y=0$, $\lambda \in \mathbb C$, are
in $L^2((a,\infty))$ (resp.\ $L^2((-\infty,a))$) for some, and,
hence, for all $a\in \mathbb R$. The differential expression
$\tau_{4n}$ is called in limit point at $\infty$ (resp.\ at
$-\infty$), if it is not in limit circle at $\infty$ (resp.\ at
$-\infty$), cf.\ \cite[Section 13.3]{W03} or \cite[Chapter 7]{Z05}.
In this case there exists one solution of $\tau_{4n}(y) - \lambda
y=0$ which is not in $L^2((a,\infty))$ (resp.\ $L^2((-\infty,a))$).

\begin{thm}\label{Erstes}
 The differential expression $\tau_{4n}$ is
in the limit point case at $\infty$ and at $-\infty$. The operator
$H$ with domain $\dom H =\cD_{\mmm}$ is self-adjoint in the Hilbert
space $L^2(\dR)$ and the spectrum of $H$ consists of isolated simple
eigenvalues which are non negative, real and accumulating to
infinity,
$$
\sigma(H) =\sigma_p(H) =\{\lambda_1, \lambda_2, \ldots \} \subset
\dR^+.
$$
\end{thm}

{\bf Proof.}\\
By \cite[Example 7.4.2 (1)]{Z05} we have  limit point case at
$\infty$ and at $-\infty$ and the operator $H$ with the domain $\dom
H =\cD_{\mmm}$ is self-adjoint in the Hilbert space $L^2(\dR)$.
Denote by $\tau_{4n,+}$ and $\tau_{4n,-}$ the restriction of the
differential expression $\tau_{4n}$ to $\dR^+$ and $\dR^-$,
respectively. Obviously, $\tau_{4n,+}$ is in limit point case at
$\infty$,
 $\tau_{4n,-}$ is in limit point case at $-\infty$ and at the other, finite,
 end point zero  the potential
 $x\mapsto x^{4n+2}$ is integrable over every interval $(-a,0)$ and
 $(0,a)$ for $a>0$. Hence, zero is a regular end point of the
 differential expressions $\tau_{4n,+}$ and $\tau_{4n,-}$,
 respectively, cf.\ \cite[Section 13.1]{W03} or \cite[Chapters 1 and 2]{LS}.
 We set
$$
\cD_{\mmm,\pm}:= \{y \in L^2(\dR^\pm) : y,y^\prime \in
AC_{loc}(\dR^\pm), y(0)=0,
 \tau_{4n}y \in L^2(\dR^\pm)\}
$$
and define $H_{4n,\pm} y := \tau_{4n,\pm}(y)$ for  $y \in \dom
H_{4n,\pm} = \cD_{\mmm,\pm}$. It follows from \cite[Lemma 3.1.2]{LS}
that the essential spectrum of $H_{4n,\pm}$ is empty. It is easily
seen that the difference of the resolvents of $H$ and the operator
$H_{4n,+} \oplus H_{4n,-}$, considered as an operator in $L^2(\dR)=
L^2(\dR^+) \oplus L^2(\dR^-)$ with domain $\cD_{\mmm,+}\oplus
\cD_{\mmm,-}$, is a finite rank operator. Hence, the essential
spectrum of $H$ is empty, that is, the spectrum of $H$ consists of
isolated eigenvalues only. Obviously, we have $H\geq 0$. Therefore
all eigenvalues are non-negative and, as $\tau_{4n}$ is in the limit
point case at $\infty$ and at $-\infty$, all eigenvalues are simple.
\qed

\begin{thm}
We have
\begin{equation}\label{commutation}
\cT \dom H = \dom H \quad \mbox{and} \quad \cP \dom H = \dom H.
\end{equation}
Moreover $H$ commutes with $\cP$, with $\cT$ and with $\PT$. Hence
 $H$ is $\PT$ symmetric and  self-adjoint in the Krein space
$(L^2(\dR), \Skindef)$. In particular we have
 $$
(\cP H)^* = H\cP = \cP H.
$$
\end{thm}

{\bf Proof.}\\
Relation (\ref{commutation}) follows immediately from the definition
of the operators $\cP$ and $\cT$ and, hence, $H$ commutes with $\cP$
and with  $\cT$,
\begin{equation}\label{commutation2}
\cP H =  H\cP \quad \mbox{and} \quad \cT H =  H\cT.
\end{equation}
 From this we conclude
$$
 \PT Hf = H\PT f
\quad \mbox{for all } f \in \dom H
$$
and, by Lemma \ref{PTLEM}, $H$ is $\PT$ symmetric. Relation
(\ref{commutation2}), Theorem \ref{Erstes} and Lemma \ref{sa} imply
the selfadjointness of $H$ in the Krein space $(L^2(\dR),
\Skindef)$. \qed

According to Theorem \ref{Erstes} all eigenvalues of $H$ are
isolated and simple. Then, see \cite[Corollary VI.6.6]{B}, the
corresponding eigenvectors are not neutral vectors in the  Krein
space $(L^2(\dR), \Skindef)$ and we obtain the following.

\begin{thm}
All eigenvalues of  $H$ are either of positive or of negative type,
\begin{equation*}
\sigma(H) = \sigma_p(H) = \sigma_{++}(H)\cup \sigma_{--}(H).
\end{equation*}
\end{thm}

\begin{rem}
We mention that the sets $\sigma_{++}(H)$ and $\sigma_{--}(H)$ are
stable under  perturbations small in gap, we refer to
\cite{ABJT,AJT,LMM1,LMaM}.
\end{rem}

\section{Domains of $\PT$ symmetric operators in the
case $\epsilon = 4n+2$}\label{section4}

Now we discuss the case $\epsilon = 4n+2$ for some $n \in \mathbb
N$, that is, we consider $\tau_{4n+2}$ defined according to
(\ref{Foundation}) via
\begin{equation*}\label{Foundation2odd}
\tau_{4n+2} (y) (x) := - y^{\prime \prime}(x) - x^{4n+4} y(x), \quad
x\in \mathbb R.
\end{equation*}

From \cite[Example 7.4.2 (2)]{Z05}\footnote{In the formulation of
\cite[Example 7.4.1]{Z05} and, hence, in \cite[Example 7.4.2
(2)]{Z05}  a minus sign is missing.} we conclude the following.
\begin{prop}\label{Erstesodd}
 The differential expression $\tau_{4n+2}$ is
in the limit circle case at $\infty$ and at $-\infty$.
\end{prop}
Recall that $\tau_{4n+2}$ is called in limit circle at $\infty$ (at
$-\infty$) if all solutions of the equation $\tau_{4n+2}(y) -
\lambda y=0$, $\lambda \in \mathbb C$, are in $L^2((a,\infty))$
(resp.\ $L^2((-\infty,a))$) for some $a\in \mathbb R$.

Again, we consider the maximal domain, i.e.,
$$
\cD_{\mmm}:= \{y \in L^2(\dR) : y,y^\prime \in AC_{loc}(\dR),
 \tau_{4n+2}(y) \in L^2(\dR)\}.
$$
In order to study all self-adjoint operators associated with
$\tau_{4n+2}$ we need to introduce some notations. For two functions
$f,g \in AC_{loc}(\dR)$ with continuous derivative, we define
$[f,g]_x$ for $x \in \dR$ via
$$
[f,g]_x := \overline{f(x)}g^\prime(x) - \overline{f^\prime(x)}g(x).
$$
Note that if $f$ and $g$ are real valued, then $[f,g]_x$ is the
Wronskian $W(f,g)$. It is well known that the limit of $[f,g]_x$ as
$x\to \infty$ and $x\to -\infty$ exists for $f,g \in \cD_{\mmm}$,
see \cite[Satz 13.4]{W03} or \cite[p.\ 184]{Z05}. We set
$$
[f,g]_\infty := \lim_{x \to \infty}[f,g]_x \quad \mbox{and} \quad
[f,g]_{-\infty} := \lim_{x \to -\infty}[f,g]_x .
$$

\begin{lem}\label{Foundation3odd}
 There exist real valued solutions $w_1,w_2 \in \cD_{\mmm}$ of
the equation
$$
\tau_{4n+2}(y) =0
$$
such that $w_1$ is an odd and $w_2$ an even function with
$$
[w_1,w_2]_{-\infty}= [w_1,w_2]_\infty =1
$$
and
$$
[w_1,w_1]_{-\infty}= [w_1,w_1]_\infty =[w_2,w_2]_{-\infty}=
[w_2,w_2]_\infty =0.
$$
\end{lem}
{\bf Proof.}\\
With each solution $z \in \cD_{\mmm}$ of the equation
$\tau_{4n+2}(y) =0$ also the function $x\mapsto \overline{z(x)}$ is
a solution of $\tau_{4n+2}(y) =0$. Hence, by Proposition
\ref{Erstesodd}, there exist two linearly independent real valued
solutions $z_1,z_2 \in \cD_{\mmm}$ of the equation $\tau_{4n+2}(y)
=0$. Denote by $z_{1,odd}$ and $z_{1,ev}$ the odd part
 of $z_1$ and the even part of $z_1$, respectively. That is
$$
z_{1,odd}(x) := \frac{z_1(x)-z_1(-x)}{2} \quad \mbox{and} \quad
z_{1,ev}:= \frac{z_1(x)+z_1(-x)}{2} \quad x\in \dR.
$$
We have $z_1=z_{1,odd} + z_{1,ev}$. Similarly, we denote by
 $z_{2,odd}$ and $z_{2,ev}$ the odd  and even part of $z_2$.
The functions $x\mapsto z_1(-x)$ and $x \mapsto z_2(-x)$ belong to
$\cD_{\mmm}$ and are solutions of $\tau_{4n+2}(y) =0$. Hence,
$z_{1,odd}, z_{1,ev}, z_{2,odd}$ and $z_{2,ev}$ belong to
$\cD_{\mmm}$ and are real valued solutions of $\tau_{4n+2}(y) =0$.
Assume that $z_{1,odd}$ and $z_{2,odd}$ are  zero functions.
 Then $z_1, z_2$ are even functions and their derivatives
$z_1^\prime, z_2^\prime$ are odd functions. We conclude for $x\in
\dR$
\begin{eqnarray}\nonumber
[z_1,z_2]_x &=& z_1(x)z_2^\prime(x) -
z_1^\prime(x)z_2(x)\\\label{OddEven}
            &=& -z_1(-x)z_2^\prime(-x) +
 z_1^\prime(-x)z_2(-x)\\\nonumber
            &=& -[z_1,z_2]_{-x}.
\end{eqnarray}
As $z_1,z_2$ are two real valued, linearly independent solution of
$\tau_{4n+2}(y) =0$, their Wronskian $[z_1,z_2]_x$ is constant for
all $x\in \dR$ and non zero, a contradiction. Hence $z_{1,odd}$ or
 $z_{2,odd}$ is not equal to zero. For simplicity, assume that
$z_{1,odd}$ is not equal to zero. We set
$$
w_1 := z_{1,odd}.
$$
By a calculation similar to (\ref{OddEven}) we see that at least one
of the functions $z_{1,ev}$ and $z_{2,ev}$ is  non zero. Let us
assume that $z_{2,ev}$ is not identically zero. Obviously,
$z_{2,ev}$ and $w_1$ are linearly independent solutions of
$\tau_{4n+2}(y) =0$, and their Wronskian $W(w_1,z_{2,ev})$ is
constant and non zero. We set
$$
w_2 := W(w_1,z_{2,ev})^{-1}z_{2,ev}.
$$
Therefore, $[w_1,w_2]_{-\infty}= [w_1,w_2]_\infty =1$,   $w_1$ is an
odd, $w_2$ an even function and $w_1,w_2$ are  solutions from
$\cD_{\mmm}$ of the equation $\tau_{4n+2}(y) =0$. The remaining
assertion of Lemma \ref{Foundation2odd} follows from the fact that
$w_1$ and $w_2$ are real valued functions. \qed

 For simplicity we set for $f \in \cD_{\mmm}$
$$
\begin{array}{ll}
 \alpha_1(f):= [w_1,f]_{-\infty},\quad & \quad
 \alpha_2(f):= [w_2,f]_{-\infty}, \\
\beta_1(f):= [w_1,f]_{\infty}, \quad & \quad \beta_2(f):=
[w_2,f]_{\infty}.
\end{array}
$$
The next lemma describes the behaviour of the above numbers under
the operators $\cP$ and $\cT$.

\begin{lem}\label{alpha_1}
For $f \in \cD_{\mmm}$ we have
$$\begin{array}{c}
\begin{array}{ll}
 \alpha_1(\cP f)= \beta_1(f),\quad & \quad
\alpha_2(\cP f)= -\beta_2(f) , \\
\beta_1(\cP f)= \alpha_1(f), \quad & \quad \beta_2(\cP f)=
-\alpha_2(f),
\end{array}\\[0.5cm]
\begin{array}{ll}
 \alpha_1(\cT f)= \overline{\beta_1(f)},\quad & \quad
\alpha_2(\cT f)= -\overline{\beta_2(f)} , \\
\beta_1(\cT f)= \overline{\alpha_1(f)}, \quad & \quad \beta_2(\cT
f)= -\overline{\alpha_2(f)}.
\end{array}
\end{array}
$$
\end{lem}
{\bf Proof.}\\
Taking into account that $w_1$ is odd and $w_1^\prime$ is even, we
see
\begin{eqnarray*}
 \alpha_1(\cP f) &=& \lim_{x\to -\infty}
-f^\prime(-x)w_1(x) -f(-x)w_1^\prime(x)\\
 &=&
\lim_{x\to \infty} f^\prime(x)w_1(x) -f(x)w_1^\prime(x) = \beta_1(f)
\end{eqnarray*}
and $\beta_1(\cP f)= \alpha_1(\cP\cP f)= \alpha_1(f).$ Similarly, as
$w_2$ is even and $w_2^\prime$ is odd,
\begin{eqnarray*}
 \alpha_2(\cP f) &=& \lim_{x\to -\infty}
-f^\prime(-x)w_2(x) -f(-x)w_2^\prime(x)\\
 &=&
\lim_{x\to \infty} -f^\prime(x)w_2(x) +f(x)w_2^\prime(x) =
-\beta_2(f)
\end{eqnarray*}
and $\beta_2(\cP f)=-\alpha_2(\cP \cP f)= -\alpha_2(f)$. The
remaining statements of Lemma \ref{alpha_1} follow immediately from
the definition of the operator $\cT$. \qed

In the sequel we will use the functions $w_1$ and $w_2$ from Lemma
\ref{Foundation2odd} to describe all boundary conditions for
self-adjoint operators associated to the differential expression
$\tau_{4n+2}$.

The following is from \cite[p.\ 64]{W03}, \cite[III.5]{JR} see also
\cite[Chapter 10, Section 4.4]{Z05}. As usual, we will consider two
different kinds of boundary conditions: mixed and separated.

All self-adjoint operators $H_{\alpha,\beta}$ associated to the
differential expression $\tau_{4n+2}$ with separated boundary
conditions are of the following form. For $\alpha, \beta \in
[0,\pi)$ we  set
\begin{equation}\label{ui}
\dom H_{\alpha,\beta} := \left\{f\in \cD_{\mmm} : \begin{array}{l}
\alpha_1(f) \cos \alpha - \alpha_2(f) \sin \alpha =0,\\
\beta_1(f) \cos \beta - \beta_2(f) \sin \beta =0.
\end{array}\right\}.
\end{equation}
Then (cf.\ \cite[Satz 13.21]{W03} and also
\cite[Chapter 10, Section 4.5]{Z05}) the operator
$H_{\alpha,\beta}$,
\begin{equation}\label{bi}
H_{\alpha,\beta}f = \tau_{4n+2}(f) \quad \mbox{for } f \in \dom
H_{\alpha,\beta},
\end{equation}
is self-adjoint in the Hilbert space $L^2(\dR)$ and the spectrum of
$H_{\alpha,\beta}$ consists of isolated simple eigenvalues
$\lambda_n$, $n\in \mathbb N$,
$$
\sigma(H) =\sigma_p(H) =\{\lambda_1, \lambda_2, \ldots \} \subset
\dR \quad \mbox{with} \quad \sum_{n\in \mathbb N} |\lambda_n|^{-2} <
\infty.
$$

All self-adjoint operators $H_{B}$ associated to the differential
expression $\tau_{4n+2}$ with mixed boundary conditions are of the
following form. For $\phi \in [0,2\pi)$, $a,b,c,d \in \dR$ with
$ad-bc=1$ we  set
\begin{equation}\label{ui1}
B:= e^{i\phi} \left(\begin{array}{cc} a&b\\c&d
       \end{array}\right),
\end{equation}

\begin{equation}\label{ui2}
\dom H_{B} := \left\{f\in \cD_{\mmm} : \left(\begin{array}{c}
\beta_1(f)\\\beta_2(f)
\end{array}\right)
= B \left(\begin{array}{c} \alpha_1(f)\\\alpha_2(f)
\end{array}\right)
\right\}.
\end{equation}
Then  (cf., e.g., \cite[Satz 13.21]{W03}) the operator $H_{B}$,
\begin{equation}\label{ui3}
H_{B}f = \tau_{4n+2}(f) \quad \mbox{for } f \in \dom H_{B},
\end{equation}
is self-adjoint in the Hilbert space $L^2(\dR)$ and the spectrum of
$H_B$ consists of isolated eigenvalues $\lambda_n$, $n\in \mathbb
N$, with multiplicity equal or less than two,
$$
\sigma(H) =\sigma_p(H) =\{\lambda_1, \lambda_2, \ldots \} \subset
\dR \quad \mbox{with} \quad \sum_{n\in \mathbb N} |\lambda_n|^{-2} <
\infty.
$$

We now formulate the main results of this section. We start with the
case of separated boundary conditions.

\begin{thm}\label{HB}
The operator $H_{\alpha,\beta}$ defined via $(${\rm\ref{ui}}$)$ and
$(${\rm\ref{bi}}$)$ with $\alpha,\beta \in [0,\pi)$ is $\PT$
symmetric if and only if
$$
\alpha + \beta =\pi \quad \mbox{or} \quad \alpha + \beta=0.
$$
In this case, $H_{\alpha,\beta}$ commutes with $\cP$ and with $\cT$.
Hence
 $H_{\alpha,\beta}$ is  self-adjoint in the Krein space
$(L^2(\dR), \Skindef)$. In particular, all eigenvalues of
$H_{\alpha,\beta}$ are either of positive or of negative type,
\begin{equation}\label{malwieder}
\sigma(H_{\alpha,\beta}) = \sigma_p(H_{\alpha,\beta}) =
\sigma_{++}(H_{\alpha,\beta})\cup \sigma_{--}(H_{\alpha,\beta}).
\end{equation}
\end{thm}
{\bf Proof.}\\
Assume $\alpha + \beta =\pi$. If, in addition, $\alpha \ne
\frac{\pi}{2}$, then we have $\sin \beta= \sin \alpha$ and $\cos
\beta = -\cos \alpha$ and with Lemma \ref{alpha_1} we conclude for
$f\in \dom H_{\alpha,\beta}$
\begin{eqnarray*}
\alpha_1(\cP f) \cos \alpha - \alpha_2(\cP f) \sin \alpha
= -\beta_1(f) \cos \beta+\beta_2(f)\sin \beta&=&0\\
\beta_1(\cP f) \cos \beta - \beta_2(\cP f) \sin \beta = -\alpha_1(f)
\cos \alpha +\alpha_2(f)\sin \alpha &=&0.
\end{eqnarray*}
Hence, $\cP f \in \dom H_{\alpha,\beta}$. If $\alpha = \beta
=\frac{\pi}{2}$ then for $f\in \dom H_{\frac{\pi}{2},
\frac{\pi}{2}}$ we have $\alpha_2(f) = \beta_2(f)=0$ and, by Lemma
\ref{alpha_1}, $\cP f\in \dom H_{\frac{\pi}{2}, \frac{\pi}{2}}$.

Assume $\alpha + \beta=0$. Then for $f\in \dom H_{0,0}$ we have
$\alpha_1(f) = \beta_1(f)=0$ and, by Lemma \ref{alpha_1}, $\cP f\in
\dom H_{0,0}$.

Hence, if $\alpha + \beta =\pi$ or $\alpha + \beta=0$ and we have
$\cP \dom H_{\alpha,\beta} \subset\dom H_{\alpha,\beta}$. Moreover,
$\dom H_{\alpha,\beta} = \cP \cP \dom H_{\alpha,\beta} \subset \cP
\dom H_{\alpha,\beta}$, that is
$$
\cP  \dom H_{\alpha,\beta} =\dom H_{\alpha,\beta}.
$$
An easy calculation gives $H_{\alpha,\beta}\cP=\cP H_{\alpha,\beta}$
and Lemma \ref{alpha_1} gives
$$
\cT  \dom H_{\alpha,\beta} =\dom H_{\alpha,\beta}, \quad \mbox{and}
\quad \cT  H_{\alpha,\beta} =H_{\alpha,\beta}\cT.
$$
Hence
$$
 \PT H_{\alpha,\beta} f = H_{\alpha,\beta}\PT f
\quad \mbox{for all } f \in \dom H_{\alpha,\beta}.
$$
By Lemma \ref{PTLEM}, $H_{\alpha,\beta}$ is $\PT$ symmetric.
 Lemma \ref{sa}
implies the selfadjointness of $H_{\alpha,\beta}$ in the Krein space
$(L^2(\dR), \Skindef)$. Relation (\ref{malwieder}) follows from the
fact that the spectrum of $H_{\alpha,\beta}$  consists only of
isolated, simple eigenvalues and from \cite[Corollary VI.6.6]{B}.

It remains to show that  $H_{\alpha,\beta}$ is not $\PT$ symmetric
if $\alpha + \beta \ne \pi$ and $\alpha + \beta\ne0$. For this we
consider functions $y_1,y_2,z_1,z_2$ from $\cD_{\mmm}$ such that
$y_j$, $j=1,2$ equal $w_j$ on the interval $(1,\infty)$, equal zero
on the interval $(-\infty,-1)$ and the functions $z_j$, $j=1,2$
equal $w_j$ on the interval $(-\infty,-1)$ and equal zero on the
interval $(1,\infty)$. Set
$$
y:= -\cos \beta y_1 + \sin \beta y_2 - \cos\alpha z_1 + \sin \alpha
z_2.
$$
We have $y\in \cD_{\mmm}$ and, by Lemma \ref{Foundation3odd},
$$
\begin{array}{ll}
 \alpha_1(y)= \sin\alpha,\quad & \quad
 \alpha_2(y)= \cos\alpha, \\
\beta_1(y)= \sin\beta, \quad & \quad \beta_2(y)= \cos\beta.
\end{array}
$$
From this we conclude $y\in \dom H_{\alpha,\beta}$ and with Lemma
\ref{alpha_1}
\begin{eqnarray*}
\alpha_1(\cP y) \cos \alpha - \alpha_2(\cP y) \sin \alpha &=&
\beta_1(y) \cos \alpha+ \beta_2(y) \sin \alpha\\
&=&
 \sin\beta \cos \alpha+ \cos\beta\sin \alpha\\
&=& \sin(\alpha + \beta) \ne 0,
\end{eqnarray*}
as $\alpha + \beta \in (0,2\pi)$ with $\alpha + \beta\ne \pi$. Hence
$\cP y \notin \dom H_{\alpha,\beta}$ and we see with Lemma
\ref{PTLEM} that $H_{\alpha,\beta}$ is not $\PT$ symmetric.  \qed

Now we formulate a similar result for the case of mixed boundary
conditions.

\begin{thm}\label{Th5}
The operator $H_B$ defined via $(${\rm \ref{ui1}}$)$,
$(${\rm\ref{ui2}}$)$ and $(${\rm \ref{ui3}}$)$ is $\PT$ symmetric if
and only if
\begin{equation}\label{PTcond}
B = \pm\left(\begin{array}{cc} a&b\\c&a
       \end{array}\right)
\mbox{ with } a^2-bc = 1.
\end{equation}
In this case, $H_{B}$ commutes with $\cP$ and with $\cT$. Hence
 $H_{B}$ is  self-adjoint in the Krein space
$(L^2(\dR), \Skindef)$. The spectrum of $H_B$ consists only of
isolated eigenvalues with multiplicity one or two.
\end{thm}
{\bf Proof.}\\
Let $f\in \dom H_B$, i.e.\
\begin{equation}\label{PPP}
 \left(\begin{array}{c}
\beta_1(f)\\\beta_2(f)
\end{array}\right)
= e^{i\phi} \left(\begin{array}{cc} a&b\\c&d
       \end{array}\right)
\left(\begin{array}{c} \alpha_1(f)\\\alpha_2(f)
\end{array}\right),
\end{equation}
for some $\phi \in [0,2\pi)$, $a,b,c,d \in \dR$ with $ad-bc=1$.
Lemma  \ref{alpha_1} implies
$$
 \left(\begin{array}{c}
\beta_1(\cP f)\\\beta_2(\cP f)
\end{array}\right)=
\left(\begin{array}{cc} 1&0\\0&-1
       \end{array}\right)
\left(\begin{array}{c} \alpha_1(f)\\\alpha_2(f)
\end{array}\right),
\left(\begin{array}{c} \alpha_1(\cP f)\\\alpha_2(\cP f)
\end{array}\right)=
\left(\begin{array}{cc} 1&0\\0&-1
       \end{array}\right)
\left(\begin{array}{c} \beta_1(f)\\\beta_2(f)
\end{array}\right)
$$
and $\cP f$ is in $\dom H_B$ if and only if
$$
\left(\begin{array}{c} \alpha_1(f)\\\alpha_2(f)
\end{array}\right)=
e^{i\phi} \left(\begin{array}{cc} 1&0\\0&-1
       \end{array}\right)
\left(\begin{array}{cc} a&b\\c&d
       \end{array}\right)
\left(\begin{array}{cc} 1&0\\0&-1
       \end{array}\right)
 \left(\begin{array}{c}
\beta_1(f)\\\beta_2(f)
\end{array}\right).
$$
With (\ref{PPP}) we see that this is the case if and only if
$$
\left(\begin{array}{c} \alpha_1(f)\\\alpha_2(f)
\end{array}\right)=
e^{2i\phi} \left(\begin{array}{cc} 1&0\\0&-1
       \end{array}\right)
\left(\begin{array}{cc} a&b\\c&d
       \end{array}\right)
\left(\begin{array}{cc} 1&0\\0&-1
       \end{array}\right)
\left(\begin{array}{cc} a&b\\c&d
       \end{array}\right)
 \left(\begin{array}{c}
\alpha_1(f)\\\alpha_2(f)
\end{array}\right)
$$
which is equivalent to
\begin{equation}\label{eequiv}
\left(\begin{array}{c} \alpha_1(f)\\\alpha_2(f)
\end{array}\right)=
e^{2i\phi} \left(\begin{array}{cc} a^2 -bc& b(a-d)\\c(d-a)&d^2-bc
       \end{array}\right)
 \left(\begin{array}{c}
\alpha_1(f)\\\alpha_2(f)
\end{array}\right).
\end{equation}
Similar as in Theorem \ref{HB} we consider functions
$y_1,y_2,z_1,z_2$ from $\cD_{\mmm}$ such that $y_j$, $j=1,2$, equals
$w_j$ on the interval $(1,\infty)$, equals zero on the interval
$(-\infty,-1)$ and the functions $z_j$, $j=1,2$, equals $w_j$ on the
interval $(-\infty,-1)$ and equals zero on the interval
$(1,\infty)$. Set
$$
y:= -ce^{i\phi}y_1 +ae^{i\phi} y_2 + z_2, \quad \mbox{and} \quad
z:=-de^{i\phi}y_1 +be^{i\phi} y_2 + z_1
$$
We have $y,z \in \cD_{\mmm}$ and
$$
\begin{array}{ll}
 \alpha_1(y)= 1,\quad & \quad
 \beta_1(y)= ae^{i\phi}, \\
\alpha_2(y)= 0, \quad & \quad \beta_2(y)= ce^{i\phi},
\end{array} \quad
\begin{array}{ll}
 \alpha_1(z)= 0,\quad & \quad
 \beta_1(z)= be^{i\phi}, \\
\alpha_2(z)= 1, \quad & \quad \beta_2(z)= de^{i\phi}.
\end{array}
$$
Hence, $y,z \in \dom H_B$, see (\ref{PPP}). Inserting $y$ and $z$ in
(\ref{eequiv}) we see
$$
e^{2i\phi}(a^2 -bc)=1, \quad b(a-d)=0=c(d-a),\quad
e^{2i\phi}(d^2-bc)=1.
$$
For $c\ne0$ it follows $a=d$ and, from $ad-bc=1$, we obtain
$a^2-bc=1$. For $c =0$ it follows $a^2=d^2=1$ and $ad= 1$ This gives
$a=d=\pm 1$. Moreover, in both cases (i.e.\ $c\ne0$ and $c =0$)
$\phi$ is either zero or $\pi$. This shows that $\cP \dom H_B
\subset\dom H_B $ if and only if (\ref{PTcond}) holds.

Hence, if (\ref{PTcond}) does not hold, there exists $f\in \dom H_B$
with $\cP f$ is not in $\dom H_B$ and we see with Lemma \ref{PTLEM}
that $H_{B}$ is not $\PT$ symmetric.

Conversely, if (\ref{PTcond})  hold, then we have $\cP  \dom H_{B}
\subset\dom H_{B}$ and $\dom H_{B} = \cP \cP \dom H_{B} \subset \cP
\dom H_{B}$, that is $\cP  \dom H_{B} =\dom H_{B}$. An easy
calculation gives $H_{B}\cP=\cP H_{B}$ and Lemma \ref{alpha_1} gives
$$
\cT  \dom H_{B} =\dom H_{B}, \quad \mbox{and} \quad \cT  H_{B}
=H_{B}\cT,
$$
hence $\PT H_{B} f = H_{B}\PT f$ for all $f \in \dom H_{B}$. By
Lemma \ref{PTLEM}, $H_{B}$ is $\PT$ symmetric.
 Lemma \ref{sa}
implies the selfadjointness of $H_{B}$ in the Krein space
$(L^2(\dR), \Skindef)$. \qed

As mentioned above, the spectrum of $H_{B}$  consists only of
isolated  eigenvalues with multiplicity less or equal to two. We
have the following.
\begin{prop}
Let the operator $H_B$ be $\PT$ symmetric and let $\lln \in
\sigma_p(H_B)$ with {\rm dim Ker}$\,(H_B-\lln)=1$, then
\begin{equation}\label{pi1}
\lln \in \sigma_{++}(H_B) \cup \sigma_{--}(H_B).
\end{equation}
If  $\lln \in \sigma_p(H_B)$ with {\rm dim Ker}$\,(H_B-\lln)=2$,
then
\begin{equation}\label{pi}
\lln \notin \sigma_{++}(H_B) \cup \sigma_{--}(H_B).
\end{equation}
\end{prop}
{\bf Proof.}\\
Relation (\ref{pi1}) follows from the fact that isolated eigenvalues
with multiplicity one
 in the  Krein space $(L^2(\dR),
\Skindef)$  are not neutral, see \cite[Corollary VI.6.6]{B}.
 Using the reasoning in the proof of Lemma
\ref{Foundation3odd} applied to the equation $\tau_{4n+2}(y) - \lln
y=0$, we find an odd and an even eigenfunction of $H_B$
corresponding to the eigenvalue $\lln$. Then the odd eigenfunction
is a negative vector in the Krein space $(L^2(\dR), \Skindef)$ and
the even eigenfunction is a positive vector in the Krein space
$(L^2(\dR), \Skindef)$ and (\ref{pi}) holds.
 \qed

 \begin{rem}
 Let the operator $H_B$ be $\PT$ symmetric.
 It is usual in the perturbation theory in Krein spaces to consider
 spectral points of type $\pi_+$ and $\pi_-$, denoted by
 $\sigma_{\pi_+}(H_{B})$ and  $\sigma_{\pi_-}(H_{B})$, see
 \cite{AJT,J3,LMaM}. The main property of these points is that they
 are invariant under compact perturbations and perturbations small
 in norm or small in the gap metric.
We mention here only that isolated eigenvalues of finite algebraic
 multiplicity are spectral points of type $\pi_+$ and $\pi_+$. Hence
\begin{equation*}
\sigma(H_{B}) = \sigma_p(H_{B}) = \sigma_{\pi_+}(H_{B})\cup
\sigma_{\pi_-}(H_{B}).
\end{equation*}
 \end{rem}

With Theorems \ref{HB} and \ref{Th5} all self-adjoint operators
associated to the differential expression $\tau_{4n+2}$  which  give
rise to a $\PT$ symmetric operator can precisely be characterized.
We wish to emphasize the following.

\begin{cor}
If $\alpha\beta\ne 0$ and $\alpha+\beta \ne \pi$, then the operator
$H_{\alpha,\beta}$ is not $\PT$ symmetric.
\end{cor}

\begin{cor}
If $d\ne a$ or $\phi$ is not zero or $\pi$, then the operator
$H_{B}$ is not $\PT$ symmetric.
\end{cor}

\end{document}